\documentclass[aps,prd,groupedaddress,showpacs]{revtex4}
\usepackage{graphicx}
\usepackage{bm}
\usepackage{amsmath}
\usepackage{latexsym}

\input{tcilatex}

\begin{document}

\title{Entropy of quantum-corrected black holes}
\author{Jerzy Matyjasek}
\email{jurek@kft.umcs.lublin.pl, matyjase@tytan.umcs.lublin.pl}
\affiliation{Institute of Physics, 
Maria Curie-Sk\l odowska University\\
pl. Marii Curie-Sk\l odowskiej 1, 
20-031 Lublin, Poland}
\date{\today}

\begin{abstract}
The approximate renormalized one-loop effective action of the quantized
massive scalar, spinor and vector field in a large mass limit, i.e.,
the lowest order of the DeWitt-Schwinger expansion involves the coincidence
limit of the Hadamard-DeWitt coefficient $a_{3}.$ Building on this and using 
Wald's approach we shall construct the general
expression describing entropy of the spherically-symmetric static
black hole being the solution of the semi-classical field equations.
For the concrete case of the quantum-corrected Reissner-Nordstr\"{o}m black hole
this result coincides, as expected, with the entropy obtained by integration of the first
law of black hole thermodynamics with a suitable choice of the integration
constant. The case of the extremal quantum corrected black hole 
is briefly considered.
\end{abstract}
\maketitle







\section{\label{intro}Introduction}

Mathematical difficulties encountered in the attempts to construct the
renormalized stress-energy tensor and other characteristics of the
quantized fields in curved background are well known, and, except
extremely simple cases, they invalidate exact treatment of the semi-
classical Einstein field equations. On the other hand, the back
reaction programme, as it is understood today, requires knowledge of
the functional dependence of the stress-energy tensor on a wide class
of metrics. Treating the renormalized stress-energy tensor of the
quantized field as a source term of the semi-classical equations, one
can, in principle, construct the self-consistent solution to the
equations and analyze evolution of the system unless the quantum
gravity effects become dominant. It is natural, therefore, that in
order to make the back reaction calculations tractable, one has to
refer to some approximations or even try numerical techniques.

Nowadays, the literature devoted to calculation of 
$\langle T_{a}^{b}\rangle$ of the quantized fields 
in the spacetimes of black holes is vast indeed. In the Schwarzschild
geometry we have good understanding of the stress-energy tensor of the
quantized massive and massless fields in the Boulware, Unruh and
Hartle-Hawking states. Specifically, due to excellent numerical work
we have results that may be considered as
exact~\cite{Elster1,Elster2,
Howard,Jens1,Jens2,Jens3,Anderson2,Anderson3}. On the other hand,
analytical~\cite{Page,Zannias,Brown1,Brown2} and semi-
analytical~\cite{FZ2,Kocio:Hadamar96,Kocio:Unr97,Kocio:vect97,Kocio:2005,Mottola1} 
approximations have been constructed and successfully applied in
numerous physically interesting cases. Moreover, at the expense of
increasing number of numerical data required to construct the stress-
energy tensor, a few best-fit models has been
proposed~\cite{Visser,Kocio:Unruh99,Kocio:IHH99}.

Unfortunately, less is known of the observables in other geometries.
There is, however, a remarkable exception: It has been shown that for
sufficiently massive fields (i.e. when the Compton length is much
smaller than the characteristic radius of curvature, where the latter
means any characteristic length scale of the geometry) the asymptotic
expansion of the effective action in powers of $m^{-2}$ may be used.
Here $m$ is a mass of the quantized field.
It is because the nonlocal contribution to the total effective action
can be neglected and the vacuum polarization part is
local and determined by the geometry of the spacetime in question.
This is a very fortunate feature as it allows a straightforward
calculation of the approximate stress-energy tensor simply by
employing the standard relation
\begin{equation}
{\frac{2}{g^{1/2}}}{\frac{\delta }{\delta g^{ab}}}S_{q}^{\left( s\right)
}\,=\,-\langle T_{ab}^{(s)}\rangle.
                                              \label{tens}
\end{equation}
One expects, by construction,  that the result satisfactorily
approximates exact $T^{ab}$ in any geometry, provided the temporal
changes of spacetime are small and the mass of the quantized field is
sufficiently large. Such a tensor of the massive scalar, spinor and
vector fields specialized to the Ricci flat geometries  has been
calculated by Frolov and Zel'nikov~\cite{FrZ82,FrZ83,FrZ84}. On the
other hand, within the framework of the sixth-order WKB approximation,
$\langle T^{ab}\rangle _{ren}$ of the massive scalar field
propagating in a general spherically-symmetric spacetime has been 
constructed by Anderson,
Hiscock and Samuel~\cite{Anderson2}. (Slightly different method has
been adopted in~\cite{Popov}). The Frolov and Zel'nikov results have
been subsequently extended to more general geometries
in~\cite{Kocio:Massive1,Kocio:Massive2}.

The stress-energy tensor of the quantized massive fields in a large
mass limit has been used in a number of physically interesting cases.
Among them a prominent role is played by the back reaction of the
quantized field upon black hole spacetimes. Such quantum corrected
solutions have been studied from various points of
view in Refs.~\cite{Anderson4,Anderson5,Anderson6,Berej,Oleg1}, where the
corrections to the location of the event horizon, mass and temperature
as well as important issue of extreme black holes have been
considered. On the other hand, less is known about the
entropy of such systems (see, however, Ref.~\cite{Oleg}).

At one loop the black hole entropy is a sum 
\begin{equation}
{\cal S}_{BH}={\cal S}_{cl}\left[ g_{ab}\right] +{\cal S}_{q}
\left[ g_{ab}\right] ,
\end{equation}
where ${\cal S}_{cl}[g_{ab}]$ is the entropy obtained form the
gravitational action whereas ${\cal S}_{q}\left[ g_{ab}\right]$ is to
be constructed from the one loop effective action and $g_{ab}$ is the
metric tensor of  the quantum corrected black hole geometry, actually
being a self consistent solution to the semi classical Einstein field
equations. Unfortunately, complexity of the field equations
invalidates, even in simplest cases, construction of the exact
solutions. However, the nature of the contribution of the quantized
fields to the total action suggests the line of attack. 
Imposing appropriate boundary conditions one can easily devise the
perturbative approach to the problem. For example, for the
spherically-symmetric and electrically charged static black hole one
can choose the {\em exact} location of the event horizon, $r_{+},$ and
the charge, $e,$ to be related with the integration constants.
Consequently, the zeroth-order solution coincides with the Reissner-
Nordstr\"{o}m geometry characterized by the exact $r_{+}$ and $e.$

Since the stress-energy tensor of the quantized field in a large mass
limit describes the vacuum polarization effects rather than the real
excitations (the latter are exponentially small) one can forget about
its material origin and treat the one loop effective action as an
additional higher curvature (derivative) correction to the total
gravitational action. As we shall see below it is a very fortunate
feature of the theory.

One can evaluate the entropy of the quantum corrected black hole
by integrating of the first law of thermodynamics. On the other
hand one can employ the Noether charge
technique~\cite{Wald1,Iyer,Jac1,Jac2,Vis1} -- a method which seems to
be best suited not only in calculations within the framework of the
higher derivative gravity and the low-energy limit of the string theory
but also in the calculations employing the
purely local one-loop effective action of the quantized massive fields.
Technically speaking, in this approach, one has to calculate
derivatives of the effective Lagrangian with respect to the Riemann
tensor and its (symmetrized) covariant derivatives and integrate the result
over a two-dimensional section of the event horizon.

In this paper using Wald's approach we shall construct the general
expression describing entropy of the spherically-symmetric static
black hole being the solution of the semi-classical field equations.
Obtained formula can be applied in any black hole spacetime provided
the assumptions necessary to construct the one loop effective action
are satisfied and the action of the classical matter fields does not
functionally depend on the Riemann tensor and its covariant
derivatives. The structure of the effective action suggests that the
entropy, except the standard Bekenstein-Hawking term, contains also
the higher curvature
contribution~\cite{Uglum,Myers95,Solodukhin95,Fursaev94}. In
general, there will be terms coming from the renormalized quadratic
action as well as the contribution of the quantized fields. 

For the
quantum corrected Reissner-Nordstr\"{o}m black hole we shall
demonstrate that identical result, up to an integration constant, 
can be obtained from the first law of black hole thermodynamics. 
It should be observed that when using Wald's
prescription, the model considered in this paper requires the zeroth-
order solution only, whereas integration of the first law requires
also its first-order corrections.

Although the quantized massive fields propagating in the black hole
geometry has been analyzed in numerous papers (see for example~\cite{Myers95}), 
their contribution to the entropy has been ignored. Indeed, after absorbing
the divergent part of the DeWitt-Schwinger action into the quadratic
Lagrangian in the process of renormalization of the bare coupling
constants, the higher order terms have been neglected. It is justified
whenever we have no interest in the entropy of the quantized field
itself and the quantum-corrected black hole, 
i. e., if we restrict ourselves to the effectively quadratic
gravity, but otherwise is unsatisfactory. To the best knowledge of the
author it is a first attempt to calculate the entropy of such systems
in a more general setting than the Schwarzschild solution~\cite{Oleg}.

\section{\label{eff_act}Effective action}

In this paper we shall restrict ourselves to the massive
scalar, spinor and vector fields satisfying the equations
(conventions are $R_{ab}=R_{\,\,\,acb}^{c}\sim \partial _{c}\Gamma _{ab}^{c}$
, signature $-,+,+,+$)
\begin{equation}
(-\nabla _{a}\nabla ^{a}\,+\,\xi R\,+\,m^{2})\phi ^{(0)}\,=0,
\end{equation}
\begin{equation}
(\gamma ^{a}\nabla _{a}\,+\,m)\phi ^{(1/2)}\,=\,0,
\end{equation}
\begin{equation}
(\delta _{b}^{a}\nabla _{c}\nabla ^{c}\,-\,\nabla _{b}\nabla
^{a}\,-\,R_{b}^{a}\,-\,\delta _{b}^{a}m^{2})\phi ^{(1)}\,=\,0,
\end{equation}
respectively. Here $\xi $ is the curvature coupling constant, 
and $\gamma ^{a}$ are the Dirac matrices obeying standard relations 
$\gamma ^{a}\gamma ^{b}\,+\,\gamma ^{b}\gamma^{a}\,=\,2g^{ab}\hat{1}.$ 
The first order term of the renormalized effective action
of the quantized scalar, spinor and vector fields in a large mass 
limit is constructed from the (traced) coincidence limit 
of the fourth Hadmard-DeWitt coefficients $a^{(s)}_{3},$ 
and can be written as~\cite{Avramidi:1986mj,Avramidi:1989ik,Avramidi:1991je}
\begin{equation}
    S_{q}^{(s)}\,=\, {1\over 32 \pi^{2} m^{2}}\int d^{4}x\, g^{1/2} 
    \left\{\begin{array}{l}
    [a^{(0)}_{3} ]\\
    -tr [a^{(1/2)}_{3}]\\
    tr [a^{(1)}_{3}]\,-\,[a^{(0)}_{3|\xi = 0}]
\end{array}\right.
\end{equation}
The coefficients $a^{(s)}_{0},$ $a^{(s)}_{1}$ and $a^{(s)}_{2}$ contribute to
the divergent part of the action, 
\begin{equation}
S_{div} = \int d^{4}x \,g^{1/2}\left(\Lambda_{B}+\frac{1}{16\pi G_{B}} R+
\alpha_{B} R^{2}+\beta_{B} R_{ab}R^{ab} + \gamma_{B} R_{abcd} R^{abcd} \right),
\end{equation}
where the subscript $B$ indicates that the constants are bare,
and have to be absorbed by the quadratic gravitational action. 
Henceforth, the renormalized Newton constant, $G,$ is set to 1. 

The total action of the system is  
\begin{equation}
S=S_{G}+S_{sq}+S_{m}+ S_{q}^{\left( s\right) }
=\int d^{4}x\,g^{1/2}\mathcal{L} ,
\label{dzial}
\end{equation}
where $S_{G}$ is the Einstein-Hilbert action 
\begin{equation}
S_{G}=\frac{1}{16\pi }\int d^{4}x\,g^{1/2}R ,
\end{equation}
$S_{sq}$ is the renormalized quadratic action
\begin{equation}
S_{sq} = \int d^{4}x \,g^{1/2}\left(\Lambda + \alpha R^{2} + \beta R_{ab}R^{ab}
+ \gamma R_{abcd} R^{abcd}\right),
\label{sq}
\end{equation}
$S_{m}$ is the action of the classical matter, and finally
$S_{q}^{\left( s\right) }$ denote the action of the quantized massive fields.

Although our primary aim is to construct the general 
expression describing the entropy of the static and spherically 
symmetric quantum corrected black holes, we use obtained results
in a spacetime of the electrically charged black hole. Here we shall
restrict ourselves to the simplest model with
\begin{equation}
S_{m} = S_{em}=-\frac{1}{16\pi }\int d^{4}x \, g^{1/2} F_{ab}F^{ab},
\end{equation}
where $ F_{ab} =\nabla_{a}A_{b}-\nabla_{b}A_{a}$ and $A_{a}$ is the
electromagnetic potential. In doing so we shall ignore other (possible) 
higher order terms such as $\left(F_{ab} F^{ab} \right)^{2}$ and 
$R_{ab} F^{ac}F_{c}^{\ b}.$

Let us return  to the effective action of the quantized fields.
It has been demonstrated that the approximate
one-loop effective action in a large mass limit is given 
by~\cite{Avramidi:1986mj,Avramidi:1989ik,Avramidi:1991je} 
\begin{align}
S_{q}^{\left( s\right) }& =\,{\frac{1}{192\pi ^{2}m^{2}}}\int 
d^{4}x\,g^{1/2}
\left( \alpha_{1}^{(s)}R\nabla _{a}\nabla ^{a}R\,+
\,\alpha _{2}^{(s)}R_{ab}\nabla
_{c}\nabla ^{c}R^{ab}\,+\,\alpha _{3}^{(s)}R^{3}\right.  
\notag \\
& +\,\alpha_{4}^{(s)}RR_{ab}R^{ab}\,
+\,\alpha _{5}^{(s)}RR_{abcd}R^{abcd}\,+
\,\alpha_{6}^{(s)}R_{b}^{a}R_{c}^{b}R_{a}^{c}\,+
\,\alpha_{7}^{(s)}R^{ab}R_{cd}R_{~a~\,b}^{c~d}  \notag \\
&\left. +\alpha_{8}^{(s)}R_{ab}R_{\,\,\,cde}^{a}R^{bcde}\,
+\,\alpha _{9}^{(s)}R{_{ab}}^{cd}R{_{cd}}^{ef}R{_{ef}}^{ab}\,
+\,\alpha_{10}^{(s)}R_{~a~b}^{c~d}R_{~e~f}^{a~b}R_{~c~d}^{e~f}\right),
                                             \label{acc}
\end{align}
where the numerical coefficients $\alpha _{i}^{(s)}$ depending on the
spin of the massive field are tabulated in Table I. It should be
emphasized that the status of the renormalized constants $\alpha,$
$\beta$ and $\gamma$ on the one
hand and coefficients $\alpha _{i}^{(s)}$ on the other is different.
Indeed, the former should be measured empirically, whereas the latter are
unambiguously determined by the spin of the field. 
All we can say at the moment is that they
are extremely small since otherwise they would give rise to
observational effects. Henceforth, for simplicity, we shall equate
them to zero and only briefly discuss their contribution to the
entropy in Sec~\ref{Noeth}.

\begin{table}[h]
\caption{The coefficients $\protect\alpha_{i}^{(s)}$ for the massive scalar,
spinor, and vector field}
\label{table1}
\begin{tabular}{cccc}
& s = 0 & s = 1/2 & s = 1 \\ 
$\alpha^{(s)}_{1} $ & ${\frac{1}{2}}\xi^{2}\,-\,{\frac{1}{5}} \xi $\,+\,${
\frac{1}{56}}$ & $- {\frac{3}{140}}$ & $- {\frac{27}{280}}$ \\ 
$\alpha^{(s)}_{2}$ & ${\frac{1}{140}}$ & ${\frac{1}{14}}$ & ${\frac{9 }{28}}$
\\ 
$\alpha^{(s)}_{3}$ & $\left( {\frac{1}{6}} - \xi\right)^{3}$ & ${\frac{1}{432
}}$ & $- {\frac{5}{72}}$ \\ 
$\alpha^{(s)}_{4}$ & $- {\frac{1}{30}}\left( {\frac{1}{6}} - \xi\right) $ & $
- {\frac{1}{90}}$ & ${\frac{31}{60}}$ \\ 
$\alpha^{(s)}_{5}$ & ${\frac{1}{30}}\left( {\frac{1}{6}} - \xi\right)$ & $- {
\frac{7}{720}}$ & $- {\frac{1}{10}}$ \\ 
$\alpha^{(s)}_{6}$ & $- {\frac{8}{945}} $ & $- {\frac{25 }{376}}$ & $- {
\frac{52}{63}}$ \\ 
$\alpha^{(s)}_{7}$ & ${\frac{2 }{315}}$ & ${\frac{47}{630}}$ & $- {\frac{19}{
105}} $ \\ 
$\alpha^{(s)}_{8}$ & ${\frac{1}{1260}}$ & ${\frac{19}{630}} $ & ${\frac{61}{
140}} $ \\ 
$\alpha^{(s)}_{9}$ & ${\frac{17}{7560}}$ & ${\frac{29}{3780}}$ & $- {\frac{67
}{2520}}$ \\ 
$\alpha^{(s)}_{10}$ & $- {\frac{1}{270}}$ & $- {\frac{1}{54}} $ & ${\frac{1}{
18}}$%
\end{tabular}
\end{table}

\section{\label{quant}The quantum corrected Reissner-Nordstr\"{o}m black hole}

Setting the renormalized constants to zero and differentiating functionally 
$S$ with respect to the metric tensor  one obtains the semi-classical
field equations in their simplest form:
\begin{equation}
G_{ab} =8\pi \left(T^{(m)}_{ab}+\langle T^{(s)}_{ab}\rangle \right),
                                                     \label{semi}
\end{equation}
where $T^{(m)}_{ab}$ and $\langle T^{(s)}_{ab}\rangle$ 
are the classical and quantum part of the stress-energy tensor, respectively.
In this section we shall briefly discuss the perturbative solution
to the semi-classical Einstein field equations describing
spherically-symmetric and electrically charged static black hole.
Although some of the results presented in this section are not new:  
the massive scalars have been considered in Refs.~\cite{Anderson4,Berej}, 
whereas solutions for the spinor and vector fields have been constructed 
in~\cite{Berej}, we shall repeat, for readers' convenience, 
the main points of their derivation. 
In doing so we shall display most of the final results in a more general
form than it was done in~\cite{Anderson4,Berej}. 

The stress-energy tensor of the massive scalar field with arbitrary 
curvature coupling has been calculated using two different methods.
The calculations presented in Ref.~\cite{Anderson2} were based on the sixth-order WKB
approximation of the solutions of the radial scalar field equation
and summation thus obtained mode functions by means of the Abel-Plana
formula.  On the other hand, to construct the stress-energy tensor 
of Ref.~\cite{Kocio:Massive1,Kocio:Massive2}
one has to functionally differentiate the effective action  with respect to the 
metric tensor. The equality of the final results is not surprising as there
is a one-to-one correspondence between the order of the WKB approximation and the
order of DeWitt-Schwinger expansion.

The approximation of the stress-energy tensor considered in this paper
is increasingly accurate as the ratio $\lambda_{C}/L$ approaches zero,
where $\lambda_{C} $ is the Compton length of the field whereas $L$
is the  characteristic radius of curvature of the black hole geometry,
i. e. when $m M >>1, $ where $M$ is the black hole mass. Detailed numerical 
calculation carried out by Taylor et al.~\cite{Anderson2,Anderson4} 
in the Reissner-Nordstr\"{o}m
background shows that there are a good agreement between the numerical 
results and the approximate $\langle T^{(s)}_{ab}\rangle .$ 
For example, for $m M \geq 2$ the 
deviation of the approximate stress-energy tensor from the 
exact one lies within a few percent.  

Since the terms constructed from  $R^{2},$ $R_{ab}R^{ab}$ and the Kretschmann
scalar are absent in the renormalized action, 
we will ignore effects of the quadratic gravity.
Their influence upon electrically charged 
black hole has been extensively studied in a number of papers
(see for example~\cite{Whitt:1985,Lousto1,Lousto3,Tryn} 
and the references cited therein)
and appropriate effects can easily be incorporated into the final result.
 
As is well known the spherically-symmetric and static 
configuration can be described  by a general line element of the form
\begin{equation}
ds^{2}=-e^{2\psi \left( r\right) }f\left( r\right) dt^{2}+f^{-1}\left(
r\right) dr^{2}+r^{2}\left( d\theta ^{2}+\sin ^{2}\theta d\phi ^{2}\right) ,
\label{le}
\end{equation}
where $f\left( r\right) $ and $\psi \left( r\right) $ are two unknown
functions. Now, let us look more closely at each of the terms in Eq.~(\ref{semi}).  
First, observe that making use of the definition, after some
algebra, one obtains the stress-energy tensor of the quantized massive field
that consists of approximately 100 terms constructed from the
curvature, its contractions and covariant derivatives. As the final
result of the calculations is rather lengthy we shall not display it
here and refer interested reader to~\cite{Kocio:Massive1,Kocio:Massive2}. 
On the other hand, the classical part of  the total stress-energy tensor,
identified here with the  electromagnetic stress-energy tensor, 
$T_{a}^{(em)b},$ and compatible with the assumed symmetry is simply 
\begin{equation}
    T_{t}^{(em)t}=T_{r}^{(em)r}=-T_{\theta }^{(em)\theta }=
      -T_{\phi }^{(em)\phi }=-\frac{C_{1}^{2}}{8\pi r^{4}},  
                                       \label{em_tens}
\end{equation}
where $C_{1}$ is to be identified with the electric charge, $e.$

Even with the simplifying substitution 
\begin{equation}
f\left( r\right) =1-\frac{2M\left( r\right) }{r},
\label{ff}
\end{equation}
the equations of motion constructed  for a line element (\ref{le})  
are too complicated to be solved exactly. Fortunately, one can easily
devise the perturbative approach to the problem, treating the higher
derivative terms (one loop effective action) as small perturbations. 
Such a procedure also guarantees exclusion of the spurious solutions
which are likely to appear as the resulting equations involve sixth-order
derivatives of the unknown functions $M(r)$ and $\psi(r)$.

Now, in order to simplify calculations and to keep control of the order 
of terms in complicated series expansions, we shall introduce another 
(dimensionless) parameter $\varepsilon ,$ substituting 
$\alpha _{i}^{\left( s\right)}\rightarrow \varepsilon 
\alpha _{i}^{\left( s\right) }$. We shall put $
\varepsilon =1$ in the final stage of calculations. For the unknown
functions $M(r)$ and $\psi (r)$ we assume that they can be expanded as 
\begin{equation}
M(r)=\sum_{i=0}^{m}\varepsilon ^{i}M_{i}(r)+O(\varepsilon ^{m+1})
\end{equation}
and 
\begin{equation}
\psi (r)=\sum_{i=1}^{m}\varepsilon ^{i}\psi _{i}(r)+O(\varepsilon ^{m+1}).
\label{psssi}
\end{equation}
The system of differential equations for $M_{i}(r)$ and $\psi _{i}(r)$ 
is to be
supplemented with the appropriate, physically motivated boundary conditions.
First, it seems natural to demand 
\begin{equation}
M(r_{+})=\frac{r_{+}^{d-1}}{2},  \label{first_type}  
\end{equation}
or, equivalently, $M_{0}\left( r_{+}\right) =r_{+}^{d-1}/2$ and $M_{i}\left(
r_{+}\right) =0$ for $i\geq 1$, where $r_{+}$ denotes the exact location of
the event horizon. Such a choice leads naturally to the horizon defined
mass. On the other hand, one can use the total mass of the system as seen by
a distant observer 
\begin{equation}
\mathcal{M}=M\left( \infty \right) .  \label{distant_mass}
\end{equation}
For the function $\psi (r)$ we shall always adopt the natural condition $%
\psi (\infty )=0$. Since the results obtained for each set of boundary
conditions are not independent, one can easily transform solution of the
first type into the solution of the second type (and vice versa). In the
course of the calculations one can safely use each of them and the particular 
choice of representation is dictated by its usefulness.

It can easily be shown that the solution parametrized by the exact 
location of the event horizon of the quantum-corrected black hole, 
$r_{+}$, and the electric charge, $e$, can be written as 
\begin{equation}
f\left( r\right) =1-\frac{r_{+}}{r}+\frac{e^{2}}{r^{2}}-\frac{e^{2}}{rr_{+}}+%
\frac{8\pi \varepsilon }{r}\int_{r_{+}}^{r}dr^{\prime }r^{\prime
}{}^{2}\langle T_{t}^{\left( s\right) t}\rangle  \label{rozw_f}
\end{equation}
and 
\begin{equation}
\psi \left( r\right) =\,4\pi \varepsilon \int_{\infty }^{r}r^{\prime }\left(
\langle T_{r}^{\left( s\right) r}\rangle - 
\langle T_{t}^{\left( s\right) t}\rangle\right) \left( 1-\frac{%
2M\left( r^{\prime }\right) }{r^{\prime }}\right) ^{-1}dr^{\prime }.
\label{rozw_psi}
\end{equation}
The zeroth-order line element is obtained by putting (formally) $\varepsilon
=0$ in $f\left( r\right) $ and $\psi \left( r\right). $  It coincides with
the Reissner-Nordstr\"{o}m solution, as expected. 

Before we proceed further, let us observe that the difference between the 
$(rr)$ and $(tt)$ components of the stress-energy tensor factorizes as 
\begin{equation}
\langle T_{r}^{(s)r}\rangle - \langle T_{t}^{(s)t}\rangle\,
=\,\left( 1-\frac{r_{+}}{r}+\frac{e^{2}}{r^{2}}-%
\frac{e^{2}}{rr_{+}}\right) F^{\left( s\right) }(r),
\end{equation}
where $F^{\left( s\right) }(r)$ is a regular function, and, consequently, the
integral (\ref{rozw_psi}) simplifies to 
\begin{equation}
\psi \left( r\right) =\varepsilon \psi _{(1)}=4\pi \varepsilon \int_{\infty
}^{r}F^{\left( s\right) }(r^{\prime })r^{\prime }dr^{\prime }.
\end{equation}
Now, inserting the zeroth-order line element into the stress-energy tensor
and performing the necessary integration one obtains the desired solution
of the semiclassical equations.
The general solution valid for any set of numerical coefficients $\alpha_{i}^{(s)}$
is too lengthy to be displayed here. Thus, we shall collect the concrete form
of functions $f\left( r\right) $ and $\psi \left( r\right)$ calculated for the scalar, 
spinor and vector field in Appendix.

It is possible to express the functions $f$ and $\psi $ in a more familiar
form by introducing the horizon defined mass $M_{H},$ i. e., to represent
the solution in terms of ($e$, $M_{H}$) rather than ($e$, $r_{+}$) or ($e$, $ {\cal M}$). This can
be easily done employing the equality 
\begin{equation}
M_{H}=\frac{r_{+}}{2}+\frac{e^{2}}{2r_{+}},  \label{mh}
\end{equation}
and with such a choice of the representation the exact location of the event
horizon is related to the horizon defined mass by the classical formula 
\begin{equation}
r_{+}=M_{H}+(M_{H}^{2}-e^{2})^{1/2}.  \label{rp}
\end{equation}

\section{\label{Entr}Entropy of the quantum corrected black hole}

\subsection{\label{Noeth}Noether charge technique}

For the Lagrangian involving the Riemann tensor and its symmetric
derivatives up some finite order $n,$ the Wald's Noether charge entropy may
be compactly written in the form \cite{Iyer,Jac1} 
\begin{equation}
{\cal S}=-2\pi \int d^{2}x\left( h\right) ^{1/2}\sum_{m=0}^{n}\left( -1\right)
^{m}\nabla _{(e_{1}...}\nabla _{e_{m})}Z^{e_{1}...e_{m};abcd}\epsilon
_{ab}\epsilon _{cd}  \label{wentr}
\end{equation}
where 
\begin{equation}
Z^{e_{1}...e_{m};abcd}=\frac{\partial \mathcal{L}}{\partial \nabla
_{(e_{1}...}\nabla _{e_{m})}R_{abcd}},  \label{wentr1}
\end{equation}
$h$ is the determinant of the induced metric, $\epsilon _{ab}$ is the
binormal to the bifurcation sphere, and the integration is carried out
across the bifurcation  surface. Actually $S$ can be evaluated not only on
the bifurcation surface but on an arbitrary cross-section of the Killing
horizon. Since $\epsilon _{ab}\epsilon _{cd}=\hat{g}_{ad}\hat{g}_{bc}-\hat{g}
_{ac}\hat{g}_{bd}$, where $\hat{g}_{ac}$ is the metric in the subspace
normal to cross section on which the entropy is calculated, one can rewrite
Eq. (\ref{wentr}) in the form 
\begin{equation}
{\cal S}=4\pi \int d^{2}x\,h^{1/2}\sum_{m=0}^{n}\left( -1\right) ^{m}\nabla
_{(e_{1}...}\nabla _{e_{m})}Z^{e_{1}...e_{m};abcd}\hat{g}_{ac}\hat{g}_{bd}.
\label{entr_Noether}
\end{equation}
The tensor $\hat{g}_{ab}$ is related to $V^{a}=K^{a}/||K||$ ($K^{a}$ is the
timelike Killing vector) and the unit normal $n^{a}$ by the formula 
$\hat{g}_{ab}=V_{a}V_{b}+n_{a}n_{b}.$

Inspection of the total action functional shows that except $S_{m}$
all the terms in the right hand side of Eq.~(\ref{dzial}) 
contribute to the entropy. 
Moreover, as the one-loop effective Lagrangian involves
two terms which are constructed from the second covariant 
derivatives of the contractions of $R^{a}_{\ bcd}$
with respect to the metric tensor, one has to take $n=2$ 
in Eq.~(\ref{wentr}). This, of course, leads to additional
computational complications.

After some algebra, for a general static 
and spherically symmetric black hole, one has 
\begin{eqnarray}
{\cal S} &=&\pi r_{+}^{2}+\frac{\varepsilon r_{+}^{2}}{12m^{2}}\left\{ 2\alpha
_{1}^{\left( s\right) }\nabla _{a}\nabla ^{a}R+\alpha _{2}^{\left( s\right)
}\left( \nabla _{a}\nabla ^{a}R_{t}^{t}+\nabla _{a}\nabla
^{a}R_{r}^{r}\right) +3\alpha _{3}^{\left( s\right) }R^{2}\right.  \nonumber \\
&&+\alpha _{4}^{\left( s\right) }\left[ R\left( R_{t}^{t}+R_{r}^{r}\right)
+R_{ab}R^{ab}\right] +\alpha _{5}^{\left( s\right) }\left(
R_{abcd}R^{abcd}+4R_{tr}^{\ \ tr}\right)  \nonumber \\
&&+\frac{3}{2}\alpha _{6}^{\left( s\right) }\left[ \left( R_{t}^{t}\right)
^{2}+\left( R_{r}^{r}\right) ^{2}\right] +\alpha _{7}^{\left( s\right) }%
\left[ R_{tr}^{\ \ tr}\left( R_{t}^{t}+R_{r}^{r}\right)
+R_{t}^{t}R_{r}^{r}+2R_{\theta }^{\theta }\left( F_{t}+F_{r}\right) \right] 
\nonumber \\
&&+\left. 2\alpha _{8}^{\left( s\right) }\left[ \left( R_{tr}^{\ \
tr}\right) ^{2}+F_{t}^{2}+F_{r}^{2}+R_{tr}^{\ \ tr}\left(
R_{t}^{t}+R_{r}^{r}\right) \right] +12\alpha _{9}^{\left( s\right) }\left(
R_{tr}^{\ \ tr}\right) ^{2}+6\alpha _{10}^{\left( s\right)
}F_{t}F_{r}\right\} _{|r_{+}}, 
\nonumber \\
 \label{entr_gen}
\end{eqnarray}
where 
\begin{equation}
F_{t}=R_{t\theta }^{\ \ t\theta }=R_{t\phi }^{\ \ t\phi }  \label{f1}
\end{equation}
and 
\begin{equation}
F_{r}=R_{r\theta }^{\ \ r\theta }=R_{r\phi }^{\ \ r\phi }.  \label{f2}
\end{equation}
Note that it is a quite general result and it can be used as long as
the black hole geometry is spherically-symmetric, the stress-energy tensor 
of the classical fields is independent of the Riemann tensor and 
$\lambda_{C}/L <<1.$
Similarly, the contribution of the quadratic part of the action to the entropy,
mostly ignored in this paper, is given by
\begin{equation}
{\cal S}_{2} = 32\pi^2 r_{+}^2\left[\alpha R
+ \frac{1}{2}\beta\left(R_{t}^{t} + R_{r}^{r}\right) 
+ 2\gamma R_{tr}^{\ \ tr}
\right]_{|r_{+}}
\end{equation}

Now, we are in position to employ the general formula describing the entropy 
of the quantum corrected black hole in the concrete case 
of the Reissner-Nordstr\"{o}m geometry. 
Inspection of Eq.~(\ref{entr_gen}) shows that it suffices to retain only the
zeroth-order solution. It is because the first term in the right hand side of
Eq.~(\ref{entr_gen}) depends solely on the radius of the event horizon, which is considered
as the exact quantity here. Consequently, inclusion of the first-order terms in the 
line element term would give rise to 
$O(\varepsilon^{2})$ terms in the final result.   
Simple calculations yield
\begin{eqnarray}
{\cal S} &=&\pi r_{+}^{2}-\frac{\varepsilon }{m^{2}r_{+}^{6}}\left[ \frac{2}{3}{%
\,e^{2}\,(e}^{2}{-r_{+}^{2})\,\alpha _{2}^{(s)}-}\frac{1}{3}e{^{4}{\alpha }%
_{4}^{\left( s\right) }}-\frac{1}{3}{(3\,r_{+}^{4}-6\,e^{2}\,r_{+}^{2}+5%
\,e^{4})\,\alpha _{5}^{(s)}}\right.  \notag \\
&&-\frac{1}{4}e{^{4}{\alpha }_{6}^{\left( s\right) }}-\frac{1}{12}e{%
^{2}\,(4\,r_{+}^{2}-7\,e^{2})\,{\alpha }_{7}^{\left( s\right) }}-\frac{1}{12}%
(14\,e^{2}\,r_{+}^{2}-3\,r_{+}^{4}-17\,e^{4})\,{\alpha }_{8}^{\left(
s\right) }  \notag \\
&&-\left. {(r_{+}^{2}-2\,e^{2})^{2}\,{\alpha }_{9}^{\left( s\right) }}+\frac{%
1}{8}{(e-r_{+})^{2}\,(e+r_{+})^{2}\,{\alpha }_{10}^{\left( s\right) }}\right]
\equiv \pi r_{+}^{2}+\Delta S^{\left( s\right) }.  \label{entr_any}
\end{eqnarray}
Substituting tabulated values of the coefficients ${\alpha
_{i}^{(s)}}$ into the above equation, one obtains 
for the scalar, spinor and vector fields
\begin{equation}
\Delta {\cal S}^{\left( 0\right) }=\frac{1}{7560m^{2}r_{+}^{2}}\left(
15r_{+}^{4}+504r_{+}^{2}e^{2}\eta -48r_{+}^{2}e^{2}-336e^{4}\eta
+49e^{4}-252r_{+}^{4}\eta \right) ,  \label{c1}
\end{equation}
\begin{equation}
\Delta {\cal S}^{\left( 1/2\right) }=\frac{1}{5040m^{2}r_{+}^{2}}\left(
77e^{4}-48r_{+}^{2}e^{2}+8r_{+}^{4}\right)  \label{c2}
\end{equation}
and 
\begin{equation}
\Delta {\cal S}^{\left( 1\right) }=\frac{1}{2520m^{2}r_{+}^{2}}\left(
148r_{+}^{2}e^{2}-7e^{4}-27r_{+}^{4}\right) ,  \label{c3}
\end{equation}
where $\eta = \xi-1/6.$
Note, that depending on the values of $q=|e|/r_{+}$ 
and the curvature coupling constant $\xi,$ the contribution of
$\Delta {\cal S}^{(s)}$ can be negative but the total entropy of the system
is, of course, always positive.
For example, for vanishing electric charge $\Delta {\cal S}^{(s)}$  
is always positive for the spinor and negative for the vector fields.
On the other hand, the sign of the contribution of the scalar field is negative
for $\xi > 19/84.$ Thus, it is positive for the conformal and the minimal
coupling.

The renormalized action of quadratic gravity~(\ref{sq})
leads to 
\begin{equation}
{\cal S}_{2} = \pi r_{+}^2-32\pi^{2} \beta \frac{e^{2}}{r_{+}^{2}} + 
64\pi^{2}\gamma \left(1-2 \frac{e^{2}}{r_{+}^{2}}\right).
\end{equation}
The higher order terms in ${\cal S}_{2}$
are 4$\pi$ times that of Ref~\cite{Myers95}, as expected. 
Since the Gauss-Bonnet invariant 
\begin{equation}
S_{GB} = \int d^{4}x g^{1/2}\left(R^{2}-4 R_{ab} R^{ab}+R_{abcd}R^{abcd}\right)
\end{equation}
has zero functional derivative with respect to the metric
tensor, the Kretschmann scalar can be relegated from the action~(\ref{sq}).
Making use of the easy-to-prove identity
\begin{equation}
 r_{+}^{2}\left( \frac{1}{2} R -  R_{t}^{t} 
- R_{r}^{r} + R_{t r}^{\ \ t r} \right)_{|r_{+}} = 1
\end{equation}
valid for the zeroth-order line element (\ref{le})
with (\ref{ff}-\ref{psssi}),
one concludes that the contribution of the $S_{GB}$ to the 
entropy is a constant independent of $r_{+}.$

Putting $e=0$ in Eqs.~(\ref{c1}-\ref{c3}) one obtains the entropy of the
quantum corrected Schwarzschild black hole. The entropy of such a black hole
confined in a spherical box of a radius $R_{0}$ has been constructed in  
Ref.~\cite{Oleg}.  It could be shown that for $R_{0}\rightarrow \infty $
the results of \cite{Oleg} coincide with $e=0$ limit of Eqs.~(\ref{c1}-\ref{c3}).

\subsection{First law}

Since the calculations of the previous subsection are rather complicated,
it is reasonable to rederive the results (\ref{entr_any}-\ref{c3}) 
using different approach. Here we shall demonstrate 
that technically independent calculation of the entropy can be carried out employing 
the first law of thermodynamics 
\begin{mathletters}
\begin{equation}
M=T d{\cal S}+\sum_{i}\mu _{i}dQ_{i},  \label{term1}
\end{equation}
where $T$ is the temperature and $\mu _{i}$ are the chemical 
potentials corresponding to the conserved
charges $Q_{i}.$ Making use of Eq. (\ref{term1}) one has 
\end{mathletters}
\begin{align}
{\cal S}& =\int T^{-1}d\mathcal{M}+{\cal S}_{0}  \notag \\
& =\int T^{-1}\left( \frac{\partial \mathcal{M}}{\partial r_{+}}\right)
_{Q_{i}}dr_{+}+{\cal S}_{0}.  \label{entr_int}
\end{align}
The integration constant ${\cal S}_{0}$ does not depend on $r_{+},$ but
possibly depends on the coupling constants.

In the present approach  it is necessary to retain 
in the line element all the terms proportional 
to $\varepsilon .$ Specifically, to construct the entropy one has to know
to the required order
both $\mathcal{M,}$ i. e., the total mass of the system 
as seen by a distant observer and the temperature.
The former quantity may be calculated from the definition (\ref{distant_mass}) and 
the formulas collected in Appendix, whereas
the temperature can be constructed using 
the Euclidean form of the line element
obtained from the Wick rotation $(t\rightarrow -it).$ 
Now, the geometry has no conical singularity as $r\rightarrow r_{+}$, provided the
`time' coordinate is periodic with a period $\beta $ given by  
\begin{equation}
\beta \,=\,4\pi \lim_{r\rightarrow r_{+}}\left( g_{tt}g_{rr}\right)
^{1/2}\left( \frac{d}{dr}g_{tt}\right) ^{-1}.  \label{beta}
\end{equation}
The Hawking temperature, $T_{H},$ is  related to $\beta $ by means of
the standard formula 
\begin{equation}
\beta \,=\,\frac{1}{T_{H}}.
\end{equation}

It can be demonstrated that the
total mass $\mathcal{M}^{\left( s\right) }$ calculated from (\ref
{distant_mass}) is given by~~\cite{Anderson4,Berej} 
\begin{eqnarray}
\mathcal{M}^{\left( s\right) } &=&M_{H}+\frac{\varepsilon }{\pi
m^{2}r_{+}^{9}}\left\{\frac{\alpha _{2}^{\left( s\right) }}{1512}e^{2}\left(
217e^{4}+252r_{+}^{4}-459e^{2}r_{+}^{2}\right) \right.  \nonumber \\
&&+\left( r_{+}^{2}-e^{2}\right) \left[ \frac{\alpha _{4}^{\left( s\right) }
}{12}e^{4}+\frac{\alpha _{5}^{\left( s\right) }}{12}\left(
3r_{+}^{4}-6e^{2}r_{+}^{2}+5e^{2}\right) +\frac{\alpha _{6}^{\left( s\right)
}}{16}e^{4}\right] \nonumber \\
&&+\frac{\alpha _{7}^{\left( s\right) }}{216}e^{2}\left(
6r_{+}^{2}-7e^{2}\right) \left( 3r_{+}^{2}-4e^{2}\right) +\frac{\alpha
_{8}^{\left( s\right) }}{432}\left(
27r_{+}^{6}+261e^{4}r_{+}^{2}-153e^{2}r_{+}^{4}-139e^{6}\right)   \nonumber \\
&&+ \frac{\alpha _{9}^{\left( s\right) }}{504}\left(
105r_{+}^{6}+909e^{4}r_{+}^{2}-585e^{2}r_{+}^{4}-445e^{6}\right) \nonumber \\
&&+ \left.\frac{
\alpha _{10}^{\left( s\right) }}{2016}\left(
21r_{+}^{6}+117e^{4}r_{+}^{2}-99e^{2}r_{+}^{4}-43e^{6}\right) \right\} .\nonumber
\\
\label{distant_M}
\end{eqnarray}
Further, restricting the general expression (\ref{beta}) to the line element (\ref
{le}) one obtains 
\begin{equation}
T_{H}=\frac{1}{4\pi }\,e^{\psi \left( r_{+}\right) }\frac{df}{dr}_{|r=r_{+}},
\label{Th}
\end{equation}
and consequently for the quantum-corrected Reissner-Nordstr\"{o}m black hole
the final result is given by 
\begin{eqnarray}
T_{H} &=&\frac{1}{4\pi r_{+}}\left( 1-\frac{e^{2}}{r_{+}^{2}}\right) +\frac{
\varepsilon }{\pi ^{2}m^{2}r_{+}^{11}}\left\{ \frac{\alpha _{2}^{\left(
s\right) }}{48}\left( r_{+}^{2}-e^{2}\right) \left( 7e^{2}-4r_{+}^{2}\right)
\right.   \nonumber \\
&&-\left( r_{+}^{2}-3e^{2}\right) \left[ \frac{\alpha _{4}^{\left( s\right) }
}{24}e^{4}+\frac{\alpha _{5}^{\left( s\right) }}{24}\left(
3r_{+}^{4}-6e^{2}r_{+}^{2}+5e^{2}\right) +\frac{\alpha _{6}^{\left( s\right)
}}{32}e^{4}\right]  \nonumber \\
&&+\frac{\alpha _{7}^{\left( s\right) }}{48}e^{2}\left(
2r_{+}^{4}+7e^{4}-6e^{2}r_{+}^{2}\right) -\frac{\alpha _{8}^{\left( s\right)
}}{96}\left( 3r_{+}^{6}+45e^{4}r_{+}^{2}-23e^{2}r_{+}^{4}-37e^{6}\right)  
\nonumber \\
&&\left. -\frac{\alpha _{9}^{\left( s\right) }}{112}\left(
7r_{+}^{6}+147e^{4}r_{+}^{2}-73e^{2}r_{+}^{4}-109e^{6}\right) +\frac{\alpha
_{10}^{\left( s\right) }}{448}\left( r_{+}^{2}-e^{2}\right) \left(
7r_{+}^{4}-8e^{2}r_{+}^{2}-e^{6}\right) \right\}.\nonumber\\
\label{te}
\end{eqnarray}

Now we are in position to calculate the entropy.
Substituting (\ref{te}) into (\ref{entr_int}), 
expanding and collecting the terms with
the like powers of $\varepsilon,$ and, finally, linearizing  the thus
obtained result,
ofter some algebra, one gets
\begin{equation}
{\cal S}_{T}={\cal S}+{\cal S}_{0}
\end{equation}
where  ${\cal S}$ is given by~(\ref{entr_any}) and ${\cal S}_{0}$
is the integration constant. Thus, ${\cal S}_{T}$ coincides with
the entropy calculated within the Wald approach provided ${\cal S}_{0}=0.$
Technically speaking, both calculations are quite different and 
the identity of the results may be considered as an important consistency check.

\subsection{The $r_{+} = r_{-}$ limit}
The issue of the entropy of the extreme black hole has been a subject
of long-standing debate. In general, there are two main lines of reasoning, yielding,
unfortunately, different results. The first one, originated in  
~\cite{Hawking}, consists in  the observation that the Euclidean topologies 
of extreme and nonextreme black holes are different. This fact has profound consequences,
the most important of which is the observation that the entropy 
of the extreme black holes does not obey area law. 
Actually, the authors of Ref.~\cite{Hawking} argued that it is zero, although
one can invent modification of the method adopted in~\cite{Hawking} to draw quite the
opposite conclusion~\cite{Mitra}.
This behaviour can be described as extremalization after quantization as opposed 
to the approach in which the order of the operations is reversed.

On the other hand, there is still growing evidence,
that the entropy of the extreme black holes, at least for some classes 
of them,  should obey the area law plus (possible) additional theory-dependent terms.
For example, in the influential paper~\cite{Vafa1}, this result has been shown  by counting 
microstates of the certain class of black holes in string theory.
Moreover, it has been explicitly demonstrated (see for example~\cite{Wit1,Wit2})
that the macroscopic entropy  calculated with the aid of 
the Wald's prescription is in agreement with the entropy obtained by
counting microstates for extremal black holes considered in Refs.~\cite{Maldacena,Vafa2}.

In the following we shall assume that the entropy formula can be extrapolated 
to the case of the  extremal black holes. 
However, even if it turns out to be wrong 
and the entropy is discontinuous, such calculations make sense.
It should be noted that adjusting the set of  parameters suitably
one can approach the extremal configuration arbitrarily close.
The entropy of such configurations
can be calculated with the aid of the Noether charge technique. Therefore, 
analyzing ${\cal S}$ as the horizons become closer and closer each other 
and eventually  merge, one can learn about the tendency of changes
and calculate the entropy of nonextremal black holes in the extremality limit.

The radial coordinates of the event and inner horizons
of the Reissner-Nordstr\"{o}m geometry are related in a simple way
\begin{equation}
r_{+} r_{-} =e^{2}.
\label{eqq}
\end{equation}
When this two horizons merge, one has a degenerate (extreme) configuration
with $r_{+} = r_{-} = |e|.$
It should be noted, however, that in the quantum corrected case Eq.~(\ref{eqq})  
is no longer valid.
Indeed, although the zeroth-order equation gives the exact location of the 
event horizon the same is not true for its second root, say, $r_{c}.$
The condition $r_{+} =r_{-}$ may be treated as a constraint equation
which can be used to relate the electric charge and the exact location of
the degenerate horizon.
Now, assuming that the radius of the event horizon can be expanded as 
\begin{equation}
r_{+}\,=\,r_{0}\,+\,\varepsilon r_{1}\,+\,\mathcal{O}(\varepsilon ^{2}),
\label{exp}
\end{equation}
where we do not ascribe any particular meaning to $r_{0}$ and $r_{1}$, one
has 
\begin{equation}
r_{+}=|e|-\frac{\varepsilon }{24\pi m^{2}\left| e\right| ^{3}}\left( 4\alpha
_{4}^{\left( s\right) }+8\alpha _{5}^{\left( s\right) }
+3\alpha _{6}^{\left( s\right) }+3\alpha _{7}^{\left(
s\right) }+6\alpha _{8}^{\left( s\right) }+12\alpha _{9}^{\left( s\right)
}\right) .  \label{reh}
\end{equation}

Similarly, one can easily explore the consequences of
vanishing of the surface gravity (temperature). Since the temperature as
given by Eq.(\ref{Th}) is defined at the event horizon, we have a system of
two equations, the first of which, $f(r_{+})=0,$ is satisfied automatically 
whereas the solution of the second one 
\begin{equation}
{\frac{1}{r_{+}}}\,-\,{\frac{e^{2}}{r_{+}^{3}}}\,+\,8\pi \varepsilon
r_{+}\langle T_{t}^{(s)t}(r_{+})\rangle\,=\,0
\end{equation}
gives the  desired result.

Now, taking the extremality limit in  the general expression (\ref{entr_any}),
after massive simplifications, one obtains 
\begin{equation}
{\cal S}=\pi e^{2}+\,\mathcal{O}(\varepsilon ^{2}).  \label{ext}
\end{equation}
This result holds for any spin of the massive field and to required order it
coincides with the analogous result calculated for the classical
Reissner-Nordstr\"{o}m black hole. Although 
the stress-energy tensor of the quantized field is known in the one loop
approximation only, it is possible to construct ${\cal S}$ of the 
nonextreme black hole up to the terms proportional to $\varepsilon ^{2}$. 
Calculations of the entropy of the extreme configuration to the
second order would require knowledge of the stress-energy tensor
beyond the one loop approximation. 

The thus obtained ${\cal S}$ is referred to as the entropy of the classical
black hole or macroscopic entropy. The former designation is somewhat 
misleading in the present context
as our black hole solution is, in fact, semiclassical. It should be noted, however,
that the effective action of the massive quantized field is constructed from curvature,
and the type of the field influences only the numerical coefficients that stand in front
of purely geometric terms in~(\ref{acc}). As such, it may be treated 
in the calculations as the classical higher derivative action functional.

 The entropy of the extremal black hole as given by Eq.~(\ref{ext}) is nonzero,
and, therefore, it contradicts the Nerst formulation of the third law of
thermodynamics, which asserts that the entropy of the system must go to zero
or a universal constant as its temperature goes to zero. However, the subtle 
point is that the Nernst formulation should not be considered as a fundamental 
law of thermodynamics. Indeed, Wald in Ref~\cite{Nernst}
constructed some explicit examples that violate the Nernst law. 

\section{Final remarks}

Now we are in position to compare our results with the results existing in
literature. The $e=0$ case has been briefly discussed 
at the end of the section~\ref{Noeth}.
On the other hand, one can treat the results of  Lu and Wise~\cite{Lu} and
$D=4$ limit of the entropy calculated in Ref.~\cite{MTT} as  special cases 
of present calculations.  
Indeed, observe that there are
similarities between $S$  of the quantized massive fields in a large mass
limit and the most general effective action involving all (time-reversal
invariant) curvature terms of dimension six considered by Lu and Wise.
Setting $\alpha _{1}^{\left( s\right) }=\alpha _{2}^{\left( s\right)
}=0,$ making use of the identity 
\begin{equation}
R_{~a~b}^{c~d}R_{~e~f}^{a~b}R_{~c~d}^{e~f}-R_{af}^{\ \ cd}R_{ce}^{\ \
ab}R_{bd}^{\ \ ef}=\frac{1}{4}R_{ab}^{\ \ cd}R_{cd}^{\ \ ef}R_{ef}^{\ \ ab}
\label{ident}
\end{equation}
in $S_{q}^{\left( s\right) },$ and, subsequently, substituting 
\begin{equation}
{\frac{1}{192\pi ^{2}m^{2}}}\alpha _{i}^{(s)}\rightarrow \alpha _{i},
\label{subs}
\end{equation}
and absorbing the right hand side of Eq.~(\ref{ident}) by the $\alpha_{9}$ term, 
one obtains precisely the action considered in Ref.~\cite{Lu} up to natural 
typographical differences. 
Lu and Wise concentrated on the influence of the sixth-order terms on the
Schwarzschild geometry. However, it is an easy exercise to generalize 
their results to the case of electrically charged black hole.  
Since it  can be easily done by equating $\alpha_{1}$ and $\alpha_{2}$ to zero and
simple rearrangement of the terms in Eq.~(\ref{entr_any}) 
we shall not display the final result here.

To construct the entropy of the quantum-corrected 
black hole making use of Eq.~(\ref{wentr})
one has to calculate the functional derivatives
of the Lagrangian with respect to the Riemann tensor and its 
symmetrized covariant derivatives. 
In this regard the (quantum-corrected) Reissner-Nordstr\"{o}m geometry
provides more sensitive test than the Schwarzschild solution.
Indeed, in the Schwarzschild geometry  
$\alpha_{1}^{(s)}R\nabla _{a}\nabla ^{a}R$ 
and
$\alpha _{2}^{(s)}R_{ab}\nabla_{c}\nabla ^{c}R^{ab}$
do not contribute to the entropy whereas in the Reissner-Nordstr\"{o}m
case one has nonvanishing contribution of $\alpha^{(s)}_{2}$ term. 
It would be interesting to analyze the black hole solutions with 
classical fields for which both terms do not vanish. 
This group of problem is under active considerations and the results
will be published elsewhere.

\appendix*
\section{}

In this appendix we collect solutions to the semi-classical equations
describing quantum-corrected Reissner-Nordstr\"{o}m black hole. 
The general solution is too lengthy to be reproduced here.
Explicit results for massive 
scalar, spinor and vector fields read~\cite{Anderson4,Berej}:
\begin{equation}
f\left( r\right) =1-\frac{r_{+}}{r}+\frac{e^{2}}{r^{2}}-\frac{e^{2}}{rr_{+}}+
\frac{8\pi \varepsilon }{m^{2}}\left( A^{\left( s\right) }\left( r\right)
+\xi B^{\left( s\right) }\left( r\right) \right) ,  \label{fr}
\end{equation}
where 
\begin{eqnarray}
A^{\left( 0\right) }(r) &=&{\frac{1153}{1960}}\,{\frac{{e}^{4}}{{r}^{8}}}+{
\frac{5}{112}}\,{\frac{{{r_{+}}}^{2}}{{r}^{6}}}+{\frac{13}{280}}\,{\frac{{e}
^{2}}{{r}^{6}}}-{\frac{1237}{30240}}\,{\frac{{{r_{+}}}^{3}}{{r}^{7}}}-{\frac{
113}{30240}}\,{\frac{1}{r{{r_{+}}}^{3}}}  \notag \\
&+&{\frac{2327}{11340}}\,{\frac{{e}^{6}}{{r}^{10}}}-{\frac{613}{1680}}\,{
\frac{{e}^{4}{r_{+}}}{{r}^{9}}}-{\frac{613}{1680}}\,{\frac{{e}^{6}}{{r}^{9}{
r_{+}}}}-{\frac{1237}{30240}}\,{\frac{{e}^{6}}{{r}^{7}{{r_{+}}}^{3}}}  \notag
\\
&+&{\frac{877}{70560}}\,{\frac{{e}^{2}}{r{{r_{+}}}^{5}}}-{\frac{1069}{70560}}
\,{\frac{{e}^{4}}{r{{r_{+}}}^{7}}}+{\frac{4169}{635040}}\,{\frac{{e}^{6}}{r{{
r_{+}}}^{9}}}-{\frac{2549}{10080}}\,{\frac{{e}^{2}{r_{+}}}{{r}^{7}}}  \notag
\\
&+&{\frac{5}{112}}\,{\frac{{e}^{4}}{{{r_{+}}}^{2}{r}^{6}}}-{\frac{2549}{10080
}}\,{\frac{{e}^{4}}{{r_{+}}\,{r}^{7}}}+{\frac{1369}{7056}}\,{\frac{{e}^{2}{{
r_{+}}}^{2}}{{r}^{8}}}+{\frac{1369}{7056}}\,{\frac{{e}^{6}}{{r}^{8}{{r_{+}}}
^{2}}}  \notag \\
&&
\end{eqnarray}
\begin{eqnarray}
A^{\left( 1/2\right) }\left( r\right) &=&{\frac{3}{280}}\,{\frac{{{r_{+}}}
^{2}}{{r}^{6}}}-{\frac{27}{140}}\,{\frac{{e}^{2}}{{r}^{6}}}+{\frac{3}{280}}\,
{\frac{{e}^{4}}{{{r_{+}}}^{2}{r}^{6}}}-{\frac{149}{15120}}\,{\frac{{{r_{+}}}
^{3}}{{r}^{7}}}+{\frac{1723}{5040}}\,{\frac{{e}^{2}{r_{+}}}{{r}^{7}}}  \notag
\\
&+&{\frac{1723}{5040}}\,{\frac{{e}^{4}}{{r_{+}}\,{r}^{7}}}-{\frac{149}{15120}
}\,{\frac{{e}^{6}}{{r}^{7}{{r_{+}}}^{3}}}-{\frac{2729}{17640}}\,{\frac{{e}
^{2}{{r_{+}}}^{2}}{{r}^{8}}}-{\frac{1073}{1764}}\,{\frac{{e}^{4}}{{r}^{8}}} 
\notag \\
&-&{\frac{2729}{17640}}\,{\frac{{e}^{6}}{{r}^{8}{{r_{+}}}^{2}}}+{\frac{2687}{
10080}}\,{\frac{{e}^{4}{r_{+}}}{{r}^{9}}}+{\frac{2687}{10080}}\,{\frac{{e}
^{6}}{{r}^{9}{r_{+}}}}-{\frac{1639}{15120}}\,{\frac{{e}^{6}}{{r}^{10}}} 
\notag \\
&+&{\frac{67}{11760}}\,{\frac{{e}^{2}}{r{{r_{+}}}^{5}}}-{\frac{13}{15120}}\,{
\frac{1}{r{{r_{+}}}^{3}}}-{\frac{767}{70560}}\,{\frac{{e}^{4}}{r{{r_{+}}}^{7}
}}+{\frac{451}{70560}}\,{\frac{{e}^{6}}{r{{r_{+}}}^{9}}}  \notag \\
&&
\end{eqnarray}
\begin{eqnarray}
A^{\left( 1\right) }\left( r\right) &=&{\frac{47849}{10080}}\,{\frac{{e}
^{2}r_{+}}{{r}^{7}}}+{\frac{47849}{10080}}\,{\frac{{e}^{4}}{r_{+}\,{r}^{7}}}-
{\frac{577}{280}}\,{\frac{{e}^{2}}{{r}^{6}}}-{\frac{37}{560}}\,{\frac{
r_{+}^{2}}{{r}^{6}}}-{\frac{37}{560}}\,{\frac{{e}^{4}}{r_{+}^{2}{r}^{6}}} 
\notag \\
&+&{\frac{611}{10080}}\,{\frac{{\mathit{r_{+}}}^{3}}{{r}^{7}}}+{\frac{611}{
10080}}\,{\frac{{e}^{6}}{{r}^{7}{\mathit{r_{+}}}^{3}}}-{\frac{10393}{3920}}\,
{\frac{{e}^{2}{\mathit{r_{+}}}^{2}}{{r}^{8}}}-{\frac{35449}{3528}}\,{\frac{{e
}^{4}}{{r}^{8}}}  \notag \\
&-&{\frac{10393}{3920}}\,{\frac{{e}^{6}}{{r}^{8}{\mathit{r_{+}}}^{2}}}+{
\frac{26879}{5040}}\,{\frac{{e}^{4}\mathit{r_{+}}}{{r}^{9}}}+{\frac{26879}{
5040}}\,{\frac{{e}^{6}}{{r}^{9}\mathit{r_{+}}}}-{\frac{31057}{11340}}\,{
\frac{{e}^{6}}{{r}^{10}}}  \notag \\
&-&{\frac{493}{14112}}\,{\frac{{e}^{2}}{r{\mathit{r_{+}}}^{5}}}+{\frac{11}{
2016}}\,{\frac{1}{r{\mathit{r_{+}}}^{3}}}+{\frac{2393}{70560}}\,{\frac{{e}
^{4}}{r{\mathit{r_{+}}}^{7}}}-{\frac{2389}{635040}}\,{\frac{{e}^{6}}{r{
\mathit{r_{+}}}^{9}}}  \notag \\
&&
\end{eqnarray}
\begin{eqnarray}
B^{\left( 0\right) }\left( r\right) &=&{\frac{11}{60}}\,{\frac{{{r_{+}}}^{3}
}{{r}^{7}}}-\frac{1}{5}\,{\frac{{{r_{+}}}^{2}}{{r}^{6}}}-{\frac{2{e}^{2}}{5{r
}^{6}}}-{\frac{91}{90}}\,{\frac{{e}^{6}}{{r}^{10}}}+{\frac{1}{60}}\,{\frac{1
}{r{{r_{+}}}^{3}}}-{\frac{29}{9}}\,{\frac{{e}^{4}}{{r}^{8}}}+{\frac{89}{60}}
\,{\frac{{e}^{4}}{{r_{+}}\,{r}^{7}}}  \notag \\
&+&{\frac{89}{60}}\,{\frac{{e}^{2}{r_{+}}}{{r}^{7}}}+{\frac{{e}^{4}}{18r{{
r_{+}}}^{7}}}-{\frac{31}{30}}\,{\frac{{e}^{6}}{{r}^{8}{{r_{+}}}^{2}}}-{\frac{
31}{30}}\,{\frac{{e}^{2}{{r_{+}}}^{2}}{{r}^{8}}}+{\frac{11}{60}}\,{\frac{{e}
^{6}}{{r}^{7}{{r_{+}}}^{3}}}  \notag \\
&+&{\frac{113}{60}}\,{\frac{{e}^{6}}{{r}^{9}{r_{+}}}}-\,{\frac{{e}^{2}}{20r{{
r_{+}}}^{5}}}-\,{\frac{{e}^{6}}{45r{{r_{+}}}^{9}}}+{\frac{113}{60}}\,{\frac{{
e}^{4}{r_{+}}}{{r}^{9}}}-\,{\frac{{e}^{4}}{5{{r_{+}}}^{2}{r}^{6}}}  \notag \\
&&
\end{eqnarray}
and 
\begin{equation}
B^{\left( 1/2\right) }\left( r\right) =B^{\left( 1\right) }\left( r\right)
=0.
\end{equation}
 
For the function $\psi (r)$ one has
\begin{eqnarray}
\psi ^{(0)} &=&{\frac{\varepsilon }{\pi m^{2}}}\left( -{\frac{29}{1120}}\,{
\frac{{{r_{+}}}^{2}}{{r}^{6}}}-{\frac{3}{80}}\,{\frac{{e}^{2}}{{r}^{6}}}-{
\frac{29}{1120}}\,{\frac{{e}^{4}}{{{r_{+}}}^{2}{r}^{6}}}\right.  \notag \\
&+&\left. {\frac{46}{441}}\,{\frac{{e}^{2}{r_{+}}}{{r}^{7}}}+{\frac{46}{441}}
\,{\frac{{e}^{4}}{{r_{+}}\,{r}^{7}}}-{\frac{229}{1680}}\,{\frac{{e}^{4}}{{r}
^{8}}}\right)  \notag \\
&+&{\frac{\varepsilon \xi }{\pi m^{2}}}\left( {\frac{7}{60}}\,{\frac{{{r_{+}}
}^{2}}{{r}^{6}}}-{\frac{8}{15}}\,{\frac{{e}^{2}{r_{+}}}{{r}^{7}}}+{\frac{7}{
30}}\,{\frac{{e}^{2}}{{r}^{6}}}\right.  \notag \\
&-&\left.{\frac{8}{15}}\,{\frac{{e}^{4}}{{r_{+}}\,{r}^{7}}}+{\frac{13}{20}}
\,{\frac{{e}^{4}}{{r}^{8}}}+{\frac{7}{60}}\,{\frac{{e}^{4}}{{{r_{+}}}^{2}{r}
^{6}}}\right)  \label{psi-fi}
\end{eqnarray}
\begin{eqnarray}
\psi^{\left( 1/2\right) } & =&\frac{\varepsilon}{\pi m^{2}}\left( -{\frac
{11}{1680}}\,{\frac{{e}^{4}}{{{r_{+}}}^{2}{r}^{6}}}-{\frac {13}{245}}\,{
\frac{{e}^{4}}{{r_{+}}\,{r}^{7}}}+{\frac {37}{1120}}\,{\frac{{e}^{4}}{{r}^{8}
}}\right.  \notag \\
& &\left. +{\frac{7}{120}}\,{\frac{{e}^{2}}{{r}^{6}}}-{\frac {13}{245}}\,{
\frac{{e}^{2}{r_{+}}}{{r}^{7}}}-{\frac{11}{1680} }\,{\frac{{{r_{+}}}^{2}}{{r}
^{6}}}\right)
\end{eqnarray}
and
\begin{eqnarray}
\psi ^{(1)} &=&{\frac{\varepsilon }{\pi m^{2}}}\left( {\frac{131}{3360}}\,{
\frac{{e}^{4}}{{{r_{+}}}^{2}{r}^{6}\,}}-{\frac{2446}{2205}}\,{\frac{{e}^{4}}{
{r_{+}}\,{r}^{7}\,}}+{\frac{2141}{1680}}\,{\frac{{e}^{4}}{{r}^{8}\,}}\right.
\notag \\
&&\left. +{\ \frac{173}{240}}\,{\frac{{e}^{2}}{{r}^{6}\,}}-{\frac{2446}{2205}
}\,{\frac{{e}^{2}{r_{+}}}{{r}^{7}\,}}+{\frac{131}{3360}}\,{\frac{{{r_{+}}}
^{2}}{{r}^{6}}}\right) .  \label{psi-fin}
\end{eqnarray}


\end{document}